\documentclass[conference]{IEEEtran}
\usepackage{color,graphicx,amsmath,amssymb,epsfig,subfigure,cite}
\newcommand{\ve}[2]{[{#1}_{1}\;{#1}_{2}\cdots{#1}_{#2}]}

\begin{document}
\title{Dirty Paper Coding using Sign-bit Shaping  and LDPC Codes}
\author{\IEEEauthorblockN{Shilpa G, Andrew Thangaraj and Srikrishna Bhashyam}
\IEEEauthorblockA{Dept of Electrical Engg\\
Indian Institute of Technology Madras\\ 
Chennai 600036, India\\
Email: andrew,skrishna@ee.iitm.ac.in}
}

\maketitle

\begin{abstract} 
Dirty paper coding (DPC) refers to methods for pre-subtraction of known interference at the transmitter of a multiuser communication system.  There are numerous applications for DPC, including coding for broadcast channels. Recently, lattice-based coding techniques have provided several designs for DPC. In lattice-based DPC, there are two codes - a convolutional code that defines a lattice used for shaping and an error correction code used for channel coding. Several specific designs have been reported in the recent literature using convolutional and graph-based codes for capacity-approaching shaping and coding gains. In most of the reported designs, either the encoder works on a joint trellis of shaping and channel codes or the decoder requires iterations between the shaping and channel decoders. This results in high complexity of implementation. In this work, we present a lattice-based DPC scheme that provides good shaping and coding gains with moderate complexity at both the encoder and the decoder. We use a convolutional code for sign-bit shaping, and a low-density parity check (LDPC) code for channel coding. The crucial idea is the introduction of a one-codeword delay and careful parsing of the bits at the transmitter, which enable an LDPC decoder to be run first at the receiver. This provides gains without the need for iterations between the shaping and channel decoders. Simulation results confirm that at high rates the proposed DPC method performs close to capacity with moderate complexity. As an application of the proposed DPC method, we show a design for superposition coding that provides rates better than time-sharing over a Gaussian broadcast channel.
\end{abstract} 
\section{Introduction} 
Situations where interference is known non-causally at the transmitter but not at the receiver model several useful multiuser communication scenarios. In \cite{1056659}, Costa introduced and studied coding for such situations and called it ``writing on dirty paper''. Dirty paper coding (DPC) is now recognized as a powerful notion central to approaching capacity on multiuser channels. 

Lattice-based ideas for DPC were suggested and shown to be capacity-approaching in \cite{1228082,1522643}. Recently, many designs of lattice-based DPC schemes have been proposed in \cite{1512417,4542781,4471934,4814366,5075904}.  Lattice-based schemes typically use cosets of a convolutional code for lattice-quantizing or shaping to minimize the energy of the difference of the coded symbols and the interfering symbols. A part of the message bits is used to choose the specific coset used in the minimization. In addition to the shaping convolutional code, an error correction code needs to be used to obtain coding gain and approach capacity. The main source of complexity in lattice-based DPC designs is combining shaping and coding encoders/decoders at the transmitter/receiver. Simple concatenation schemes are not applicable because of the following reasons - outer shaping followed by inner coding results in unshaped parity symbols that increase transmitted energy, while outer coding followed by inner shaping results in a poor inner code that needs to be iteratively decoded at the receiver with the outer code.

In \cite{4471934}, encoding is done on a combined trellis of the source code (Turbo TCQ) and a channel code (Turbo TCM). At the receiver, decoding is done for Turbo TCM followed by syndrome computation to recover message bits. The transmitter is complex in \cite{4471934} because of the use of the joint trellis. The DPC method proposed in \cite{4814366} is similar to that of \cite{4471934}. In \cite{4542781}, multilevel coding is used, and there are different codes for different bits of the symbols. At the receiver, iterations have to be performed between decoders for some of the channel codes and the shaping decoder. In \cite{5075904} and \cite{1523319}, shaping follows channel coding and the receiver performs iterations between the shaping and channel decoders.

In this work, we propose a lattice-based method that uses a novel combination of a convolutional code for sign-bit shaping and a low density parity check (LDPC) code for channel coding. As shown in specific designs and simulations, the method provides good shaping and coding gains at moderate complexity. The main idea for reducing complexity at the receiver is the introduction of a one-codeword delay at the transmitter, and the shaping of symbols from current message bits combined with parity bits from the previous codeword. This enables the LDPC decoder to be run first at the receiver (with a one-codeword delay) without any need for iterations with a shaping decoder. As an application, we use the proposed DPC method to design codes for superposition coding in two-user Gaussian broadcast channels. By simulations, we show that rate points outside the time-sharing region are achieved.

The rest of the paper is organized as follows. After a brief review of the lattice-based DPC coding method in Section \ref{CLS}, we present the proposed DPC method in Section \ref{PS}. This is followed by description and simulation of specific designs of DPC codes in Section \ref{SR}. In Section \ref{BC}, design of a superposition scheme using the proposed DPC method is described and simulation results are presented. Concluding remarks are made in Section \ref{conc}.   
\section{Lattice Dirty Paper Codes}
\label{CLS}
In a Gaussian dirty paper channel, the received symbol vector $\mathbf{Y}=[Y_1\;Y_2\cdots Y_n]$ is modeled as 
$$\mathbf{Y}=\mathbf{X}+\mathbf{S}+\mathbf{N},$$
where $\mathbf{X}=[X_1\;X_2\cdots X_n]$ denotes the transmitted vector, $\mathbf{S}=[S_1\;S_2\cdots S_n]$ denotes the interfering vector assumed to be known non-causally at the transmitter and $\mathbf{N}$ denotes the additive Gaussian noise vector. The transmit power is assumed to be upper-bounded by $\frac{1}{n}E[|\mathbf{X}|^2]\leq P_X$ per symbol, and the interference power is denoted $\frac{1}{n}E[|\mathbf{S}|^2]= P_S$ per symbol. The noise variance per symbol is denoted $P_N$. In \cite{1056659}, Costa shows that the capacity of the dirty paper channel is $\frac{1}{2}
\log\left(1+\frac{P_X}{P_N}\right)$ i.e. known interference can be canceled perfectly at the transmitter.

The interfering vector $\mathbf{S}$ is used as an input in the encoding process and plays an important role to determine a suitable transmit vector $\mathbf{X}$. A coding strategy for choosing $\mathbf{X}$ needs to overcome the imminent addition of $\mathbf{S}$ and protect the transmitted information from the addition of the noise $\mathbf{N}$. Such coding strategies are called dirty paper coding (DPC) methods. 

In \cite{1522643}, a dirty paper coding (DPC) scheme based on lattice strategies was proposed and shown to achieve the capacity of the dirty paper channel. We follow \cite{1512417} for a brief review of the transmitter and receiver structure in the lattice DPC method \cite{1522643}. Let $\Lambda$ denote an $n$-dimensional lattice with fundamental Voronoi region $\nu$ having averaged second moment $P\left(\Lambda\right)=P_X$ and normalized second moment $G\left(\Lambda\right)$. Also let $\textbf{\textit{U}}\sim \text{Unif}\left(\nu\right)$ i.e. $\textbf{\textit{U}}$ is a random variable (dither) uniformly distributed over $\nu$. The lattice transmission approach of \cite{1522643}\cite{1512417} is as follows.
\newline
\begin{itemize}
	\item \textit{Transmitter}: The input alphabet $X$ is restricted to $\nu$. For any $\textbf{\textit{v}}\in\nu$, the encoder sends
	\begin{equation}
	\textbf{\textit{X}}=\left[\textbf{\textit{v}}-\alpha\textbf{\textit{S}}-\textbf{\textit{U}}\right]\text{mod}\ \Lambda,
\end{equation}
where $\alpha=\frac{P_X}{P_X+P_N}$ is a MMSE scaling factor \cite{1512417}.
\item\textit{Receiver}: The receiver computes 
\begin{equation}
	\textbf{\textit{Y}}^{\prime}=\left[\alpha\textbf{\textit{Y}}+\textbf{\textit{U}}\right]\text{mod}\ \Lambda
\end{equation}
\end{itemize}
\noindent
The channel from \textbf{\textit{v}} to $\textbf{\textit{Y}}^{\prime}$ defined by (1) and (2) is equivalent in distribution to 
\begin{equation}
	\textbf{\textit{Y}}^{\prime}=\left[\textbf{\textit{v}}+\textbf{\textit{N}}^{\prime}\right]\text{mod}\ \Lambda{,}
\end{equation}
where 
\begin{equation}
	\textbf{\textit{N}}^{\prime}=\left[\left(1-\alpha\right)\textbf{\textit{U}}+\alpha\textbf{\textit{N}}\right]\text{mod}\ \Lambda.
\end{equation}
Lower bounds on achievable rates for the above equivalent channel is shown in \cite{1512417} to be equal to 
\begin{equation}
	I\left(\textbf{\textit{V}}{;}\textbf{\textit{Y}}^{\prime}\right)\geq\frac{1}{2}\log_{2}\left(1+\text{SNR}\right)-\frac{1}{2}\log_{2}\left(2\pi eG\left(\Lambda\right)\right).
\end{equation}
For optimal shaping, $G\left(\Lambda\right)\rightarrow\frac{1}{2\pi e}$ and we approach capacity of the dirty paper channel. Note that the dither is assumed to be known at the transmitter and receiver (say, through the use of a common seed in a random number generator).


\section{Proposed Scheme}
\label{PS}
\begin{figure*}[ht!]
\begin{center}
\includegraphics[scale = 1.3]{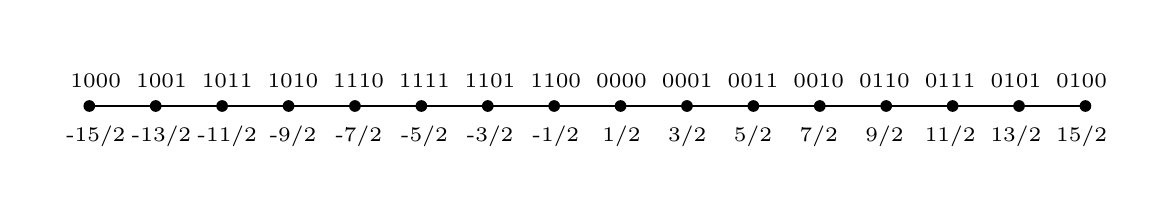}
\end{center}
\caption{16-PAM constellation.}
\label{16pam} 
\end{figure*}

The proposed scheme uses a convolutional code for sign-bit shaping \cite{Forney:1992uq} and low density parity check (LDPC) codes for channel coding. We assume a $M$-PAM signal constellation with a carefully chosen bit-to-symbol mapping that is compatible with sign-bit shaping and bit-interleaved coded modulation (BICM) \cite{Caire:1998cz}. For $M=16$, the constellation and mapping are shown in Fig. \ref{16pam}.
The mapping in Fig. \ref{16pam} is suited for sign-bit shaping, since a flip of the most significant bit results in a significant change in symbol value for all possible 4-bit inputs. Also, the mapping is mostly Gray except for a few symbol transitions. Gray mapping is known to be the most effective mapping for BICM with LDPC codes. This heuristic choice of mapping enables the possibility of good shaping and coding gains to be obtained simultaneously. As expected, larger values of $M$ will result in larger shaping gains in our design, and we stick to the 16-PAM shown in Fig. \ref{16pam} for illustration and simulation.

\subsection{Encoder Structure}
\label{enc}
\begin{figure*}[htb]
\centering
\scalebox{0.60}
{\begin{picture}(770,370)(0,0)
\put(0,220){\vector(1,0){40}}
\put(0,10){\line(1,0){90}}
\put(90,220){\vector(1,0){40}}
\put(15,265){\makebox(0,0){ $k^{\prime }$}}
\put(15,250){\makebox(0,0){ Message }}
\put(15,235){\makebox(0,0){ Bits }}
\put(40,195){\framebox(50,50){{${H^{-1}}^{T}$}}}
\put(15,55){\makebox(0,0){ $\left(k-k^{\prime}\right)$}}
\put(15,40){\makebox(0,0){ Message }}
\put(15,25){\makebox(0,0){Bits }}
\put(130,45){\framebox(100,200){{}}}
\put(180,152.5){\makebox(0,0){ Algorithm }}
\put(180,167.5){\makebox(0,0){ Viterbi  }}
\put(180,182.5){\makebox(0,0){ Shaping Coder }}
\put(180,137.5){\makebox(0,0){ minimizes the }}
\put(180,122.5){\makebox(0,0){ energy of }}
\put(180,107.5){\makebox(0,0){ 
$\left(\mathbf{v}-\alpha \mathbf{S}\right)\ \text{mod}\ M$ }}
\put(90,10){\line(1,0){250}}
\put(230,220){\vector(1,0){60}}
\put(260,265){\makebox(0,0){ $s$}}
\put(110,265){\makebox(0,0){ $s=\frac{n}{log_2{M}}$}}
\put(110,250){\makebox(0,0){ Bits }}
\put(260,250){\makebox(0,0){ Shaped }}
\put(260,235){\makebox(0,0){ Bits }}

\put(290,45){\framebox(100,200){{}}}

\put(130,45){\framebox(100,200){{}}}
\put(340,152.5){\makebox(0,0){ Encoder }}
\put(340,167.5){\makebox(0,0){ LDPC  }}

\put(340,137.5){\makebox(0,0){ of rate }}
\put(340,122.5){\makebox(0,0){$\left(k-{k}^{\prime}+s\right)/n$}}

\put(340,10){\vector(0,1){35}}

\put(390,220){\vector(1,0){100}}
\put(390,70){\vector(1,0){250}}
\put(530,220){\vector(1,0){20}}
\put(590,185){\vector(1,0){50}}
\put(615,170){\makebox(0,0){ $\underline{a}\ \underline{b}\ \underline{c}$ }}

\put(440,205){\makebox(0,0){ $n-\left(k-k^{\prime}+s\right)$}}
\put(440,190){\makebox(0,0){ Parity }}
\put(440,175){\makebox(0,0){ Bits }}

\put(490,195){\framebox(40,50){{Delay
}}}

\put(515,140){\makebox(0,0){ 
$k-k^{\prime}$}}
\put(515,125){\makebox(0,0){ Message bits}}
\put(390,155){\vector(1,0){160}}
\put(550,150){\framebox(40,75){{$\prod$}}}

\put(515,55){\makebox(0,0){ 
$s$}}
\put(515,40){\makebox(0,0){Shaped Bits }}
\put(615,55){\makebox(0,0){$\underline{z}$ }}



\put(640,45){\framebox(100,200){{}}}
\put(690,160){\makebox(0,0){ Mapping }}
\put(690,145){\makebox(0,0){ to  }}

\put(690,130){\makebox(0,0){ $M-\text{PAM}$ }}
\put(690,115){\makebox(0,0){ $\text{map}\left(\mathbf{z}\mathbf{a}\mathbf{b}\mathbf{c}\right)$}}

\put(740,145){\vector(1,0){40}}
\put(790,145){\circle{20}}
\put(790,145){\makebox(0,0){ $\tiny{\sum}$  }}

\put(790,175){\vector(0,-1){20}}
\put(790,135){\vector(0,-1){20}}
\put(760,155){\makebox(0,0){ $\mathbf{v}$  }}
\put(790,185){\makebox(0,0){ $-\alpha \mathbf{S}$  }}
\put(790,35){\makebox(0,0){ $\left(\mathbf{v}-\alpha \mathbf{S}\right)\ \text{mod}\ M$  }}
\put(765,65){\framebox(50,50){$\mod M$}}

\put(790,65){\vector(0,-1){20}}

\put(180,305){\vector(0,-1){60}}

\put(605,305){\line(0,-1){120}}

\put(605,305){\line(-1,0){310}}
\put(295,305){\line(-1,0){115}}


\put(540,350){\makebox(0,0){ $n-s$}}

\put(540,335){\makebox(0,0){ Parity plus Message }}
\put(540,320){\makebox(0,0){ Bits }}

\end{picture}

}
\caption{Encoder structure.}
\label{encoder}
\end{figure*}
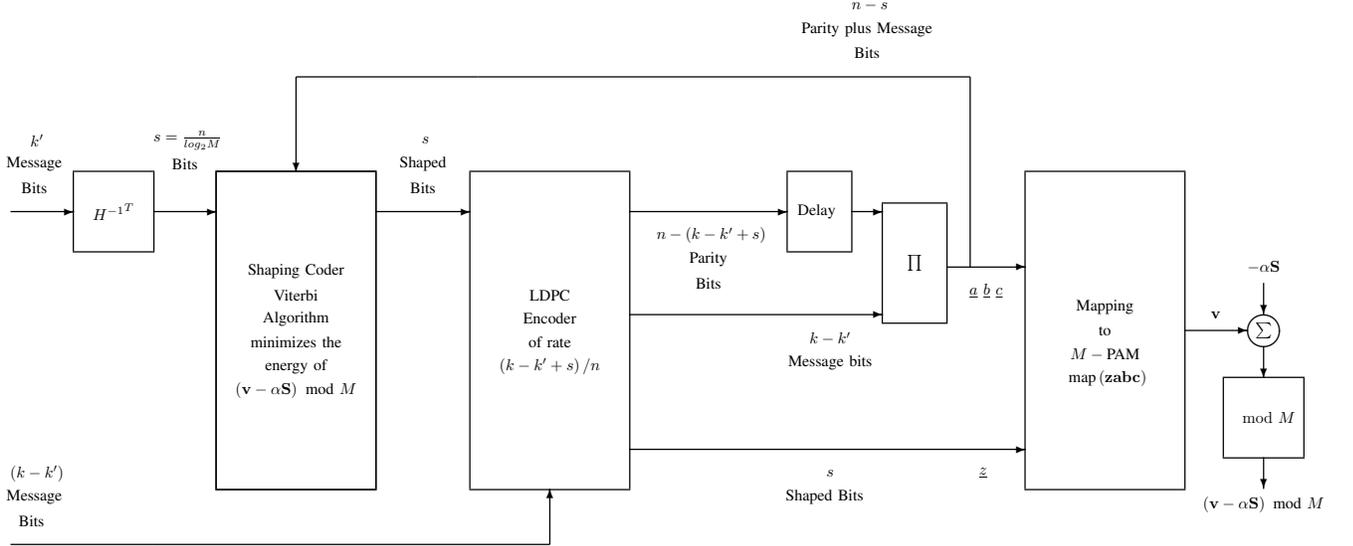

The encoder structure for the proposed scheme is as shown in Fig. \ref{encoder}. 
\thicklines
We describe the operations in the encoder at time step $T$ or in the $T$-th block. A $k$-bit message $\mathbf{m}=\ve{m}{k}$ is encoded into a $s$-symbol vector $\mathbf{u}=\ve{u}{s}$ from the $M$-PAM constellation $A=\{-(M-1)/2,\cdots,-1/2,1/2,\cdots,(M-1)/2\}$, where $s=\frac{n}{\log_2M}$ is assumed to be an integer. Let $l=\log_2M$ and let $f_{M}:\{0,1\}^{l}\rightarrow A$ denote the bit-to-symbol mapping. The bits that map to the $i$-th symbol are denoted $z_ia_{2i}a_{3i}\cdots a_{li}$; the sign-bit vector is denoted $\mathbf{z}=\ve{z}{s}$, and we define vectors $\mathbf{a}_j=\ve{a_{j}}{s}$ for $2\leq j\leq l$. Finally, we have $\mathbf{v}=f_{M}(\mathbf{z}\mathbf{a_{2}}\cdots\mathbf{a_{l}})$, where $f_M$ operates component-wise on a vector input.

Let us assume that the vectors $\mathbf{a}_{j}$, $2\leq j\leq l$ are available at the encoder. The sign-bit shaping convolutional code is used to determine the sign-bit vector $\mathbf{z}$ as follows. A part of the message $\mathbf{m'}=[m_{1}\;m_{2}\cdots m_{k'}]$ with $k'<k$ bits is mapped to a coset leader of the convolutional code using an inverse syndrome former \cite{Forney:1992uq}. Note that we need the rate of the convolutional code to be $1-k'/s$. Let the coset chosen by $\mathbf{m'}$ be denoted $C(\mathbf{m'})$. The sign-bit vector $\mathbf{z}$ is chosen from $C(\mathbf{m'})$ so as to minimize the squared sum (energy) of the vector $(\mathbf{v}-\alpha \mathbf{S})\mod M$, where $\alpha=\frac{P_X}{P_X+P_N}$ is the MMSE factor and $\mathbf{S}$ is the interference vector.  That is,
\begin{equation}
\mathbf{z}=\text{arg}\min_{\mathbf{u}\in C(\mathbf{m'})}|(f_{M}(\mathbf{u}\mathbf{a}_{2}\cdots\mathbf{a}_{l})-\alpha \mathbf{S})\mod M|^{2}.
\label{eqn:min}
\end{equation}
The minimization in (\ref{eqn:min}) is implemented using the Viterbi algorithm \cite{Forney:1992uq}. 

The $\mathbf{a}_{j}$, $2\leq j\leq l$ are determined as follows. An $(n,k-k'+s)$ LDPC code is used at the encoder with a systematic encoder $E:\{0,1\}^{k-k'+s}\rightarrow\{0,1\}^{n}$. Let $\mathbf{m}''=[\mathbf{z}\;m_{k'+1}\cdots m_{k}]$ be input to the systematic LDPC encoder to obtain the codeword $E(\mathbf{m}'')=[\mathbf{m}''\;\mathbf{p}_{T}]$, where $\mathbf{p}_{T}$ is the parity-bit vector for the $T$-th block.  The parity-bit vector is delayed by one time step. For the $T$-th block, the $n-s=s(l-1)$ bits in $[m_{k'+1}\cdots m_{k}]$ and $\mathbf{p}_{T-1}$ are rearranged by a permutation $\Pi$ to form the vectors $\mathbf{a}_{j}$, $2\leq j\leq l$. This permutation is necessary in an implementation of BICM \cite{Caire:1998cz}.

Note that both the shaping and coding objectives have been met at the encoder. The transmitted symbols $\mathbf{v}-\alpha\mathbf{S}\mod M$ have minimal energy in the lattice defined by sign-bit shaping using the convolutional code. Selected bits in successive blocks of symbols form codewords of the LDPC code. In summary, the encoder structure achieves DPC shaping and LDPC coding with bit-interleaved modulation.

\subsection{Decoder Structure}
\label{dec}
\begin{figure*}[htb]

\centering
\scalebox{0.75}
{\begin{picture}(520,170)(0,0)

\put(0,80){\vector(1,0){40}}
\put(0,95){\makebox(0,0){ $\widehat{\mathbf{Y}}=\alpha \mathbf{Y}+\mathbf{U}$ }}
\put(40,40){\framebox(60,80){{Demapper}}}
\put(100,110){\vector(1,0){170}}
\put(300,110){\vector(1,0){20}}
\put(270,85){\framebox(30,30){{$\prod^{-1}$}}}

\put(100,50){\vector(1,0){110}}
\put(210,42.5){\framebox(40,15){{Delay}}}
\put(250,50){\vector(1,0){70}}

\put(150,150){\makebox(0,0){ $n-\left(k-k^{\prime}+s\right)$}}
\put(150,135){\makebox(0,0){ LLRs of }}
\put(150,120){\makebox(0,0){ Parity Bits }}

\put(100,90){\vector(1,0){110}}
\put(210,82.5){\framebox(40,15){{Delay}}}
\put(250,90){\vector(1,0){20}}
\put(300,90){\vector(1,0){20}}

\put(150,80){\makebox(0,0){ $\left(k-k^{\prime}\right)$}}
\put(150,70){\makebox(0,0){ LLRs of Message Bits}}

\put(150,40){\makebox(0,0){ $\frac{n}{log_2{M}}$}}
\put(150,25){\makebox(0,0){ LLRs of  sign bits}}

\put(320,40){\framebox(60,80){{}}}
\put(350,90){\makebox(0,0){ LDPC }}
\put(350,70){\makebox(0,0){ Decoder }}

\put(380,55){\vector(1,0){130}}
\put(380,105){\vector(1,0){40}}
\put(470,105){\vector(1,0){40}}
\put(420,90){\framebox(50,30){{$H^{T}$}}}

\put(400,135){\makebox(0,0){ $s$ }}
\put(400,120){\makebox(0,0){ Bits }}

\put(535,120){\makebox(0,0){ $k^{\prime}$ }}
\put(535,105){\makebox(0,0){ Message }}
\put(535,90){\makebox(0,0){ Bits }}

\put(535,70){\makebox(0,0){ $k-k^{\prime}$ }}
\put(535,55){\makebox(0,0){ Message }}
\put(535,40){\makebox(0,0){ Bits }}

\end{picture}

}
\caption{Decoder structure.}
\label{decoder}
\end{figure*}
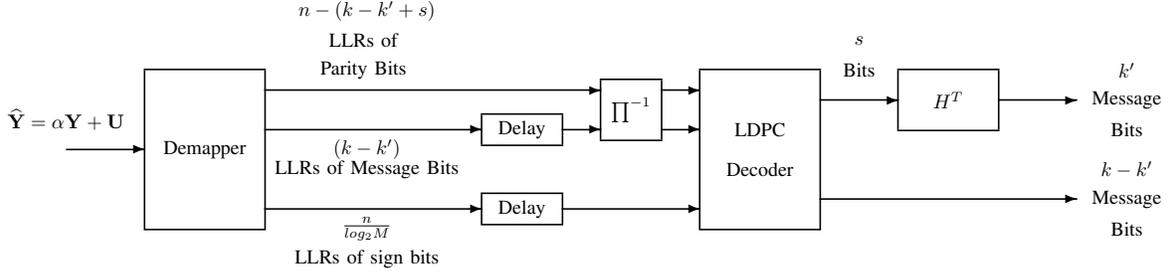

The decoder for the proposed scheme is as shown in Fig.\ref{decoder}. 
The demapper computes log likelihood ratios (LLRs) for the bits from the received symbols in $\widehat{\mathbf{Y}}=\alpha\mathbf{Y}+\mathbf{U}$. The LLRs of the $\left(k-k^{\prime}\right)$ message bits after a delay of one time step, and the LLRs of the $n-\left(k-k^{\prime}+s\right)$ parity bits are de-interleaved. The $s=\frac{n}{log_2{M}}$ LLRs of the sign bits after a delay on one time step, and the $n-s$ output LLRs of the de-interleaver are given as the input to the LDPC decoder. The LDPC decoder outputs $k-k^{\prime}$ message bits and $s$ bits of the sign bit vector of the previous block. Now, the $s$-bit sign vector is passed through the syndrome former to recover the remaining $k^{\prime}$ message bits. 

The demapper function at the receiver has to calculate LLRs taking into account the modulo $M$ operation at the encoder \cite{1512417}. Therefore, the received constellation $A_{R}$ is a replicated version of the $M$-PAM constellation $A$ used at the transmitter (assuming that scaling factors have been corrected at the receiver). That is,
$$A_{R}=\{A-rM,\cdots,A-M,A,A+M,\cdots,A+rM\}.$$
The number of replications $r$ is chosen so that the average power of $A_{R}$ is approximately equal to the total average power $P_{X}+P_{S}$, and the bit mapping of the symbol $a+jM$ $(a\in A, 1\leq j\leq r)$ is the same as that for $a$. The LLR for the $i$-th bit in the $j$-th symbol $\widehat{Y}_{j}$ is computed according to the constellation $A_{R}$ using the following formula:
\begin{equation}	
L_i=\frac{\displaystyle\sum_{a\in A_{R}:\text{bit}\;i=0}
\exp\left(-\frac{1}{2}\frac{\left(\widehat{Y}_{j}-a\right)^2}{\alpha P_N}\right)}
{\displaystyle\sum_{a\in A_{R}:\text{bit}\;i=1}
\exp\left(-\frac{1}{2}\frac{\left(\widehat{Y}_j-a\right)^2}{\alpha P_N}\right)}. \nonumber
\end{equation}
Since the constellation mapping is nearly Gray, iterations with the demapper do not provide significant improvements in coding gain, particularly for large $M$.
\section{Simulation Results}
\label{SR}
For simulations, we have taken $n=40000$, $k=30000$, $k'=5000$ with $M=16$; this results in $s=10000$. The constellation mapping is as given in Fig. \ref{16pam}. We have chosen a rate-1/2 memory 8 (256 state) non-systematic convolutional code with generator polynomials $\left(D^8+D^5+D^4+D^2+D+1,\;D^8+D^7+D^4+D^2+1\right)$ as the sign-bit shaping code. A non-systematic convolutional code is used to avoid error propagation problems.

\begin{figure}[htb]
\centering
\includegraphics[width = 3.4in]{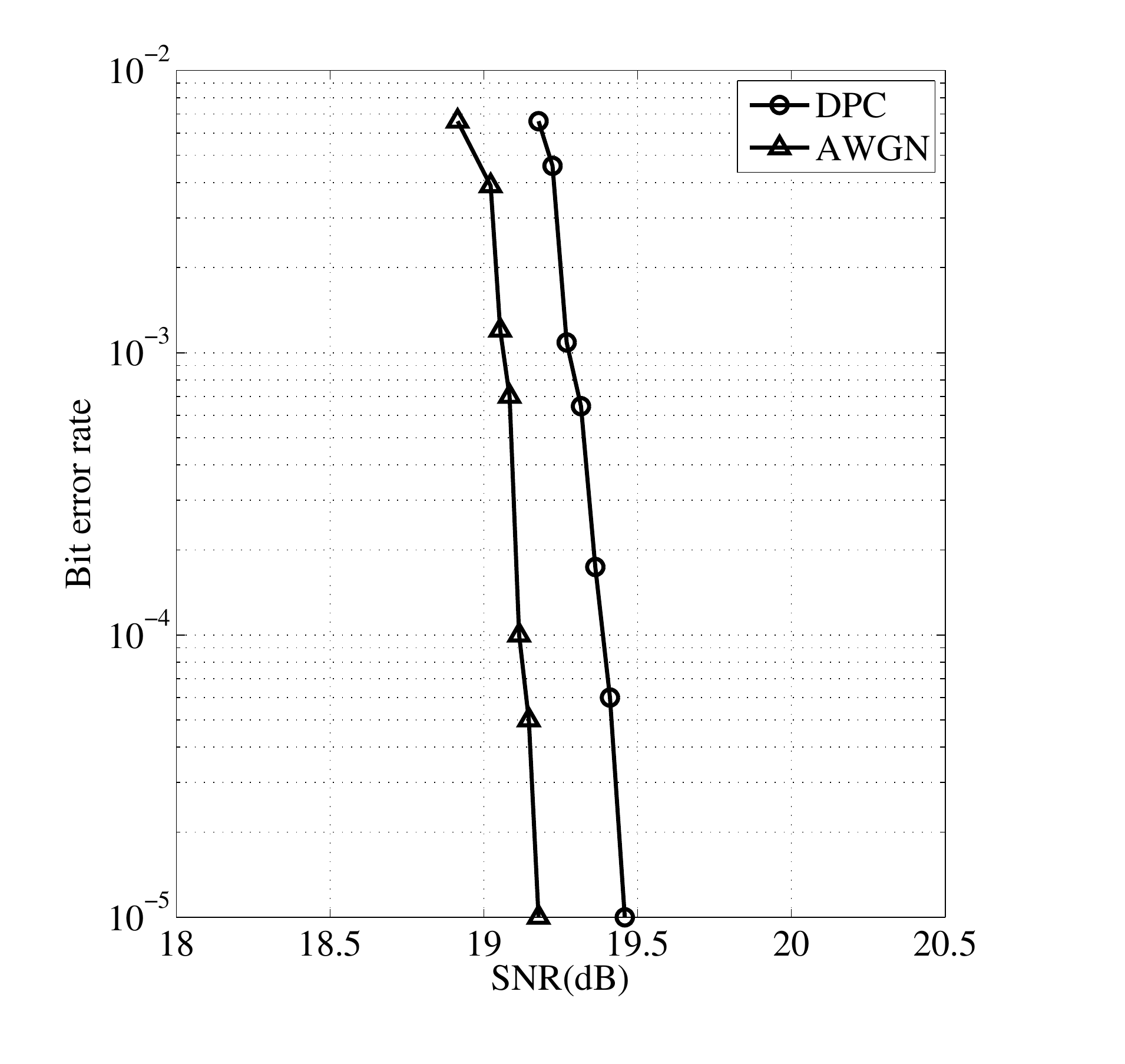}
\caption{BER plot for DPC and AWGN.}
\label{dpc} 
\end{figure}

A randomly constructed irregular LDPC code (40000, 35000) of rate $7/8$ with variable node degree distribution: $0.1256x+0.7140x^2+0.1604x^9$ and check node degree distribution $x^{31}$ is used as the channel code. The overall rate of transmission is seen to be $\frac{30000}{40000}\times 4=3$ bits per channel use. Fig. \ref{dpc} shows BER plots over an AWGN channel and a DPC channel with interference known at the transmitter. 
The interfering vector was generated at random for different power levels. The plot with interference did not change appreciably for all power levels of interference, and we have provided one plot for illustration. We see that a BER of $10^{-5}$ is achieved at a SNR of 19.45dB with interference, and at a SNR of 19.33 dB without interference. We have simulated 1000 blocks of length 40000 to obtain sufficient statistics for a BER of $10^{-5}$. 

The AWGN capacity at an SNR of $10\log_{10}(2^6-1)=17.99$ dB for a rate of 3 bits per channel use. This shows that we are 1.46 dB away from ideal dirty paper channel capacity. The granular gain $G(\Lambda)=2^{C^*}/{6S_x}$ is computed from the simulations to be 1.282dB \cite{1512417}, where $C^*=3.5$ is the rate before channel coding, and $S_x$ is the transmit power (obtained through simulations). From this, the shaping loss is calculated as follows: 
\begin{equation}
	10\text{log}_{10}{\dfrac{2\pi eG\left(\Lambda\right)2^{2C^{*}}-1}{2^{2C^{*}}-1}}=0.2548\;\text{dB}.
\end{equation}
So, of the total gap of 1.46 dB, we have a shaping gap of 0.2548dB, and a coding gap of 1.2052dB to capacity. 


\section{Application to Gaussian Broadcast Channel }
\label{BC}
\begin{figure}[htb]
\centering
\scalebox{0.65}{
\begin{picture}(350,250)(0,0)  
\put(0,20){\framebox(200,200){}}
\put(15,165){\vector(1,0){50}}
\put(35,185){\makebox(0,0){\scriptsize{User 1}}}
\put(35,175){\makebox(0,0){\scriptsize{Message bits}}}
\put(65,140){\framebox(50,50){\scriptsize{}}}
\put(90,172.5){\makebox(0,0){\scriptsize{Shaped and}}}
\put(90,157.5){\makebox(0,0){\scriptsize{LDPC coded}}}
\put(90,140){\vector(0,-1){40}}
\put(110,130){\makebox(0,0){\scriptsize{$X_1$}}}
\put(110,120){\makebox(0,0){\scriptsize{as $\mathbf{S}$ for}}}
\put(110,110){\makebox(0,0){\scriptsize{User 2}}}

\put(115,165){\line(1,0){50}}
\put(165,165){\vector(0,-1){35}}
\put(172.5,147.5){\makebox(0,0){\scriptsize{$X_1$}}}

\put(15,75){\vector(1,0){50}}
\put(35,95){\makebox(0,0){\scriptsize{User 2}}}
\put(35,85){\makebox(0,0){\scriptsize{Message bits}}}
\put(65,50){\framebox(50,50){\scriptsize{}}}
\put(90,75){\makebox(0,0){\scriptsize{DPC coded}}}
\put(115,75){\line(1,0){50}}
\put(165,75){\vector(0,1){35}}
\put(172.5,92.5){\makebox(0,0){\scriptsize{$X_2$}}}

\put(165,120){\circle{20}}
\put(165,120){\makebox(0,0){{$\sum$}}}
\put(175,120){\vector(1,0){50}}
\put(187.5,127.5){\makebox(0,0){\scriptsize{$X$}}}

\put(225,120){\vector(1,1){80}}

\put(225,120){\vector(1,-1){80}}

\put(245,180){\makebox(0,0){\scriptsize{$Y_1=X+N_1$}}}

\put(245,60){\makebox(0,0){\scriptsize{$Y_2=X+N_2$}}}

\put(305,175){\framebox(50,50){\scriptsize{$\text{Receiver 1}$}}}

\put(305,15){\framebox(50,50){\scriptsize{$\text{Receiver 2}$}}}

\end{picture}

}

\caption{Block diagram for a two-user Gaussian broadcast channel.}
\label{HBC}
\end{figure}
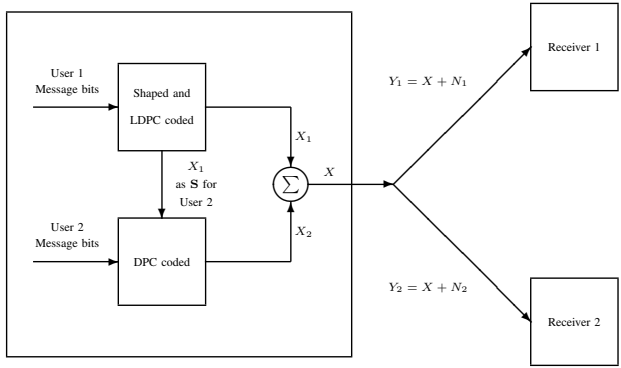

We use the proposed scheme for superposition coding in a two-user Gaussian broadcast channel $Y_1=X+N_1$ and $Y_2=X+N_2$ with $P_{N_1}>P_{N_2}$. We let $P_{X_1}=\left(1-\beta\right)P$ and $P_{X_2}=\beta P$, where $P$ is the total transmit power. Here, User 2 is coded using DPC considering User 1 as interference. User 1 is shaped using sign-bit shaping and coded using an LDPC code over $M$-PAM. Fig. \ref{HBC} shows a block diagram of the transmitter and receivers. 
The encoder structure for User 1 is as in Fig. \ref{encoder} with the interference vector $\mathbf{S}=\mathbf{0}$. Hence, for User 1, the shaping coder minimizes the energy of $\mathbf{v}$. The demapper at Receiver 1 calculates LLR for the $i$-th bit in the $j$-th receiver symbol $Y_{1j}$ using the following formula.
\begin{equation}
	L_i=\frac{\displaystyle\sum_{a\in A:\text{bit}\;i=0}\left\{p(a)\exp\left\{{-\frac{1}{2}\frac{\left(Y_{1j}-a\right)^2}{\beta P+P_{N_1}}}\right\}\right\}}{\displaystyle\sum_{a\in A:\text{bit}\;i=1}\left\{p(a)\exp\left\{{-\frac{1}{2}\frac{\left(Y_{1j}-a\right)^2}{\beta P+P_{N_1}}}\right\}\right\}}, \nonumber
\end{equation}
where $p(a)$ for $a\in A$ represents the {\it a priori} probability of the $M$-PAM symbol $a$. At the receiver, we approximate $p_i$ using a Gaussian distribution with variance $P_S$ assuming that the distribution of $M$-PAM symbols is approximately Gaussian.

\begin{figure}[htb]
\centering
\includegraphics[width = 3.5in]{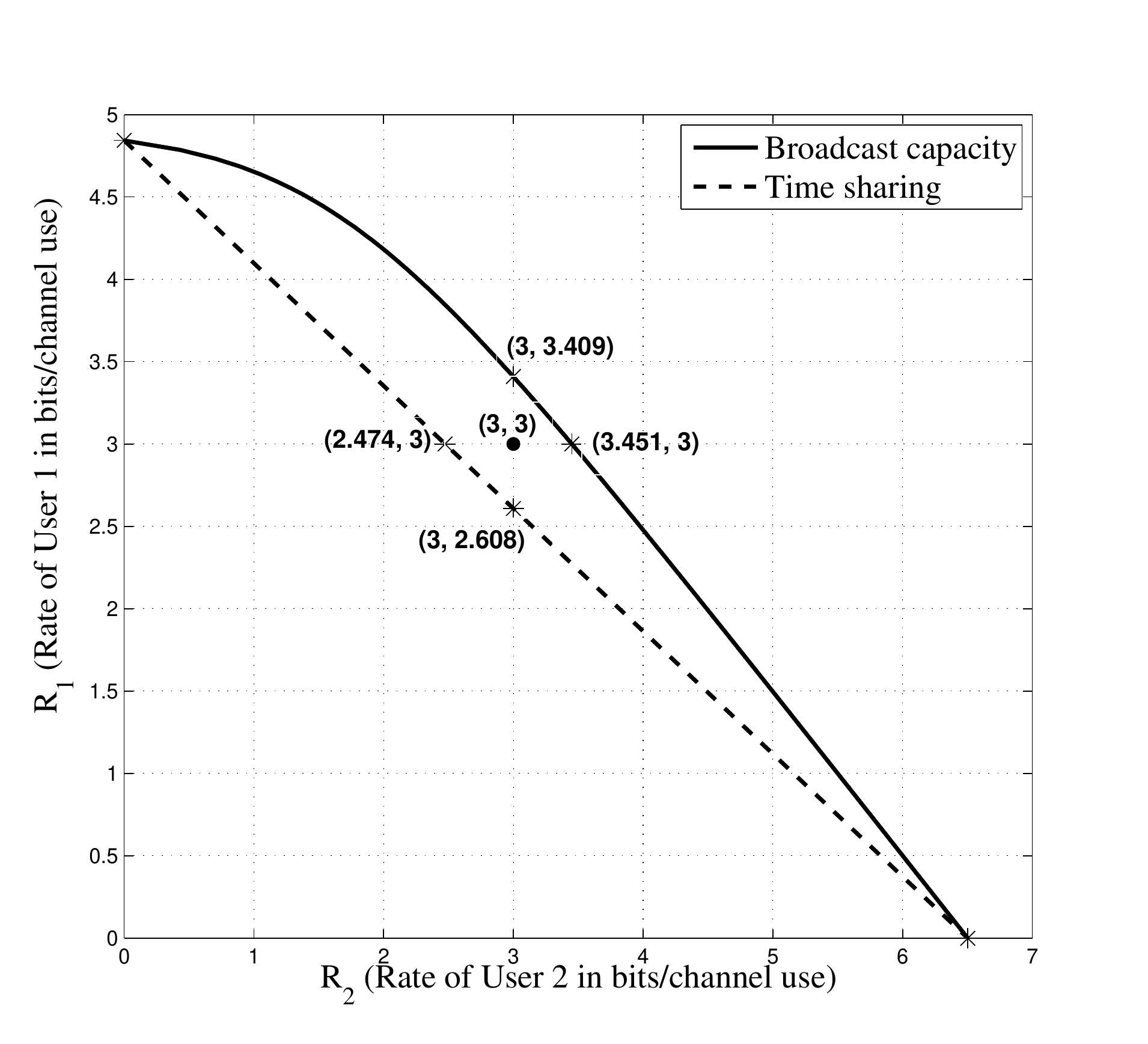}
\caption{Two-user Gaussian broadcast channel capacity region.}
\label{256state1} 
\end{figure}

We simulated a two user degraded broadcast channel with $P_{N_1}=0.9$ and $P_{N_2}= 0.09$ using the proposed scheme with parameters from Section \ref{SR}. The total transmit power, power for User 1 and power for User 2 required for a bit error rate of $10^{-5}$ (at both receivers) are estimated from the simulation and denoted $P$, $P_{x_1}$ and $P_{X_2}$, respectively. The SNR for Receiver 1 is computed as $10\log_{10}{\left(\frac{P_{X_1}}{P_{X_2}+P_{N_1}}\right)}=19.1791$ dB. Since DPC is done for User 2, the effective SNR at Receiver 2 is computed as $10\log_{10}{\left(\frac{P_{X_2}}{P_{N_2}}\right)}=19.4574$ dB. Comparison with the SNR needed for a single user capacity of 3 bits per channel use (which is 17.99 dB) shows that the total loss for both the users is about 2.4642dB. Fig. \ref{256state1} shows the (3, 3) rate pair in the capacity region of the two-user Gaussian broadcast channel with total transmit power $P$ and noise power $P_{N_1}$, $P_{N_2}$, which is defined by $R_1\leq \frac{1}{2}\log\left(1+\frac{(1-\beta)P}{\beta P+P_{N_1}}\right)$, $R_2\leq \frac{1}{2}\log\left(1+\frac{\beta P}{P_{N_2}}\right)$ for $0\leq\beta\leq1$. 
We see that the (3,3) rate point is clearly outside the time-sharing region.
%

\section{Conclusions }
\label{conc}
In this work, we have proposed a method for designing lattice-based schemes for dirty paper coding using sign-bit shaping and LDPC codes. Simulation results show that the proposed design performs 1.46dB away from the dirty paper capacity for a block length of $n=40000$ at the rate of 3 bits/channel use. This performance is comparable to other results in the literature. However, as discussed in this article, a novel method for combining shaping and coding results in good gains at lesser complexity in our design, when compared to other lattice-based strategies. As an application, we have designed a superposition coding scheme for Gaussian broadcast channels that is shown to perform better than time-sharing through simulations.

Out of the 1.46 dB gap to capacity, about 1.2 dB is gap attributed to a sub-optimal choice of the LDPC code. Optimizing the LDPC code will require use of genetic algorithms and asymmetric density evolution \cite{1542413}, which are topics for future work. 

\end{document}